\documentclass{aa}
\usepackage{graphics}

\def\e{{\rm e}}
\def\eps{\varepsilon}
\def\th{{\rm th}}
\def\cre{{\rm CRe}}
\def\crp{{\rm CRp}}
\def\ls{\,^<\!\!\!\!_\sim\,\,}

\begin{document}



\title{A Comparison of Radio and X-Ray Morphologies of Four Clusters of
Galaxies Containing Radio Halos}


\author{F. Govoni\inst{1,2}, T. A. En{\ss}lin\inst{3},
L. Feretti\inst{1}, G. Giovannini\inst{1,4}}
\offprints{fgovoni@astbo1.bo.cnr.it} \institute{ Istituto di
Radioastronomia del CNR, Via P. Gobetti 101, I-40129 Bologna, Italy
\and Dipartimento di Astronomia, Univ. di Bologna, via Ranzani 1,
I-40127, Bologna, Italy \and Max-Planck-Institut f\"{u}r Astrophysik,
Karl-Schwarzschild-Str.1, 85740 Garching, Germany \and Dipartimento di
Fisica, Univ. di Bologna, via Berti-Pichat 6/2, I-40127, Bologna,
Italy}

\maketitle

\markboth{F. Govoni et al.: A comparison of radio and X-ray morphologies of four clusters of galaxies containing radio halos}{F. Govoni et al.: A comparison of radio and X-ray morphologies of four clusters of galaxies containing radio halos}

\begin{abstract}
Clusters of galaxies may contain cluster-wide, centrally located,
diffuse radio sources, called halos. They have been found to
show morphologies similar to those of the X-ray emission. To
quantify this qualitative statement we performed a point-to-point
comparison of the radio and the X-ray emission for four
clusters of galaxies containing radio halos: Coma, Abell 2255,
Abell 2319, Abell 2744. Our study leads to a linear relation between
the radio and the X-ray surface brightness in two clusters, namely
Abell 2255 and Abell 2744. In Coma and A2319 the radio and the X-ray
brightnesses seem to be related with a sub-linear power
law. Implications of these findings within simple radio halo formation
models are briefly discussed.

\end{abstract}


\section{Introduction\label{sec:intro}}
The presence of cluster-wide magnetic fields and relativistic
electrons in cluster of galaxies is revealed by the detection of
radio halos in the center of some clusters of galaxies.  Radio halos
are diffuse, extended radio sources associated with the
intra-cluster medium (ICM) rather than with a particular galaxy.

The physical mechanism producing the population of radio-synchrotron
emitting relativistic electrons is not yet determined. However, it is
clear from cooling time arguments that such mechanism has to
operate in the ICM, in order to explain the large extensions
($\simeq$ 1Mpc) of radio halos.

In general, clusters containing radio halos are characterized by
the presence of a recent merger event, that probably
provides the energy source for the relativistic electrons, 
by the absence of strong cooling flows and by high X-ray
luminosity (Feretti 2000).

Several suggestions for the mechanism transferring energy to the
relativistic electrons have been made (see En{\ss}lin (2000) for a
recent review). These include in-situ acceleration by plasma
and by shock waves, particle injection from radio galaxies,
acceleration out of the thermal pool, secondary electrons resulting
from hadronic collisions of relativistic protons with the ICM gas
protons, and combinations of these processes.

It is, therefore, important to establish observable quantities
which allow a discrimination between these models. Important
observable quantities used so far in the literature are: the
global radio spectral index, the emission and spectral index profile,
the total radio power and its correlation to cluster parameters
such as temperature or X-ray luminosity.

It is the purpose of this paper to present a new observable
quantity: the point-to-point correlation of the radio and the
X-ray surface brightness in clusters containing radio halos.  Its
suitability is demonstrated by the striking similarities of the radio
halo morphology and the thermal X-ray emission of halo clusters.
These morphological similarities indicate an energetical relation
between the thermal gas and the relativistic plasma.  This observable
is straightforward to determine from observational data, can be
constrained by theoretical models and does not depend on any
assumption.

We present in Sect. 2 the methodological approach of the radio - X-ray
comparison.
The results are given in Sect. 3 and discussed in Sect. 4.
 
Intrinsic parameters are computed with H$_0$~=~50~km~s$^{-1}$Mpc$^{-1}$
and q$_0$~=~0.5.

\section{Data Analysis}

So far, the similarity of the radio halo and the X-ray morphology of
the clusters has been pointed out in the literature
(Deiss et al. 1997, Feretti 1999, Liang et al. 2000) but it has
never been used to constrain theoretical models.

Here we present the first quantitative analysis of the relation
between the radio and the X-ray brightness distribution.  We
carried out such analysis for the following four clusters:
Coma (Abell 1656), Abell 2255, Abell 2319, and Abell 2744, whose
properties are summarized in Table 1.

Except for A2744, we analyzed already published radio and
X-ray images. Therefore, we refer to the previous works for the
details of the radio and X-ray data.

\begin{figure}
\resizebox{9 cm}{!}{\includegraphics {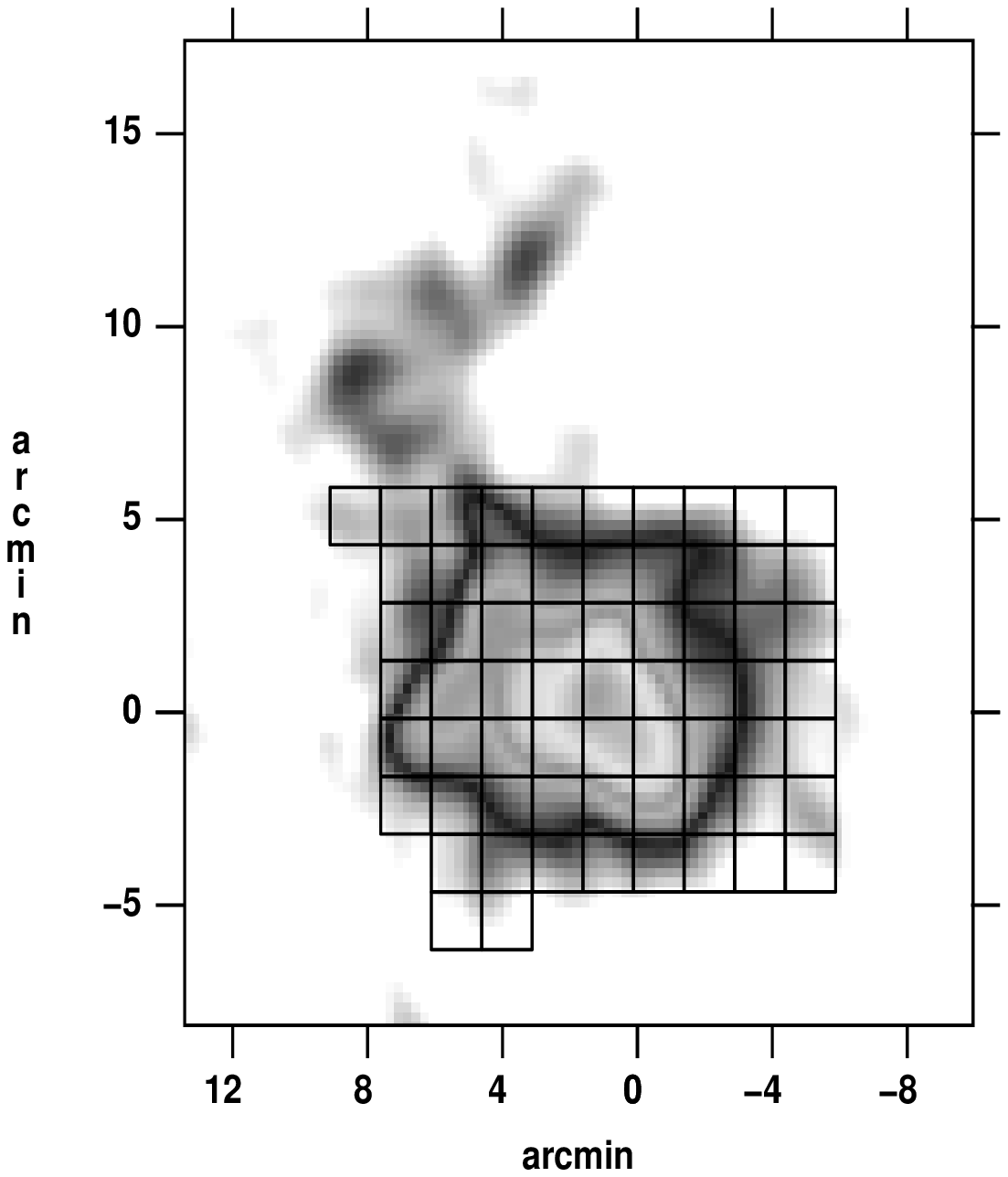}}
\resizebox{9 cm}{!}{\includegraphics {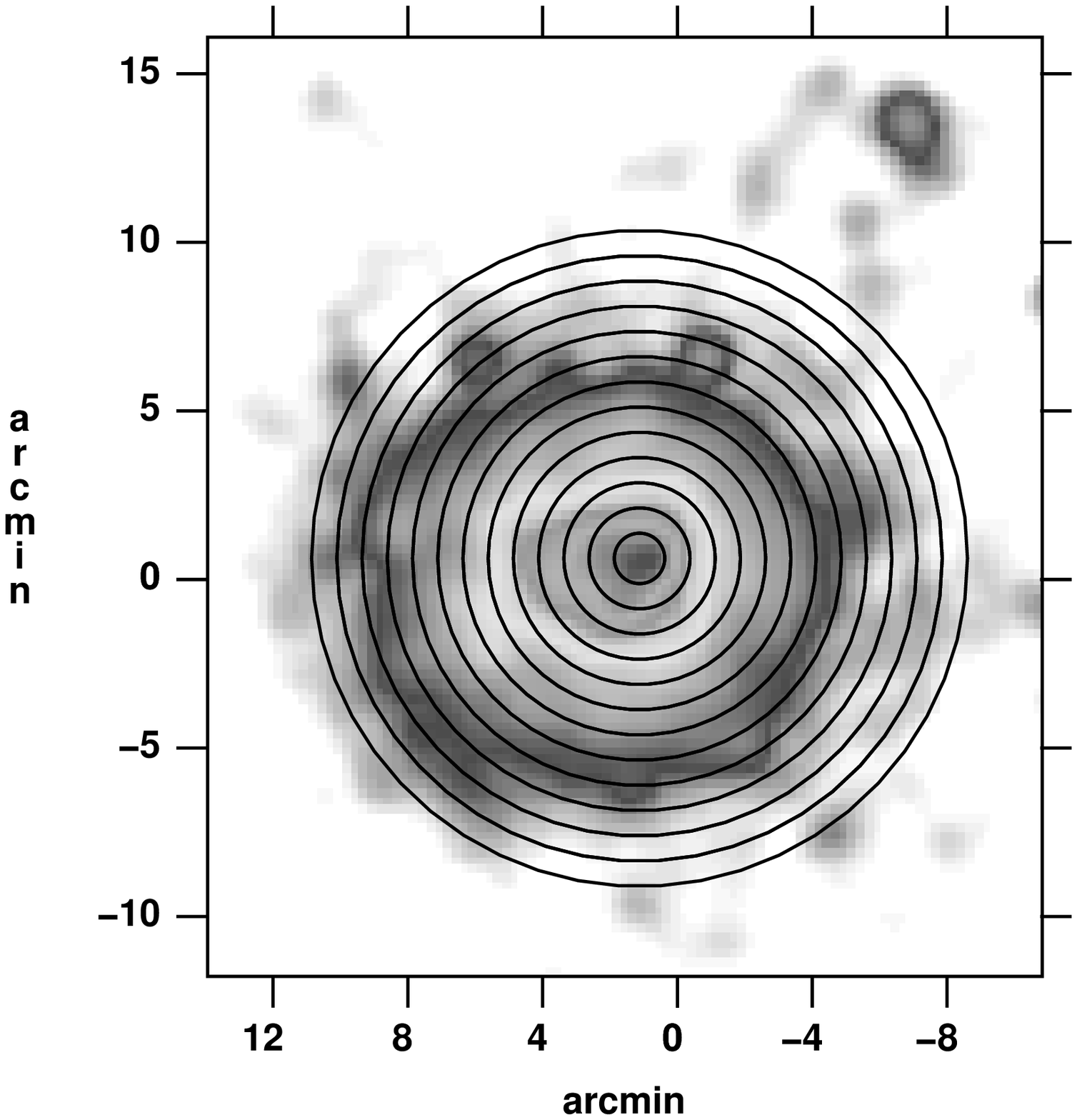}}
\hfill
\caption[]{Top panel: grid used to analyze the radio and the X-ray
images of A2255. Here, the grid is overlapped to the radio halo.
The filamentary structure extended toward the North-East 
is a relic source (Feretti et al. 1997a). 
The relic source was not included in
the grid because this source is not related to the central radio
halo under discussion here. Bottom panel: concentric rings used to
obtain the brightness profiles in A2255, superimposed to the X-ray
image.}
\label{synage}
\end{figure}

The analyzed X-ray surface brightness images were obtained in the 
0.5-2.0 keV energy band (only the Coma surface brightness image
was obtained in the 0.5-2.4 keV energy band).
The X-ray background was taken into account in the analysis of the X-ray
images.

The presence of a visible Sunyaev-Zel'dovich (SZ) effect was
established in three of our four clusters (see Table 1).  However,
the radio halo surface brightness is not corrected for a SZ
effect contamination since this is significant only
for frequencies higher than 5 GHz. The images
discussed here have been obtained at much lower frequencies (0.3
or 1.4 GHz).

\begin{table*}
\caption{Cluster parameters}
\begin{flushleft}
\begin{tabular}{ccccc}
\hline
\noalign{\smallskip}
Name       &  $^a$redshift   & $^b$L$_X$ (0.1-2.4 keV) & $^c$T    & $^d$Sunyaev-Zel'dovich \\
           &                 & 10$^{44}$erg/s          & (keV)& effect            \\
\noalign{\smallskip}
\hline
\noalign{\smallskip}
 A2255     &  0.0806        &  4.79      & 7.3            & no        \\
 Coma      &  0.0231        &  7.21      & 8.4            & yes        \\
 A2744     &  0.3080        & 22.05      & 11.0            & marginal detection \\
 A2319     &  0.0557        & 13.71      & 9.1            & yes        \\
\noalign{\smallskip}
\hline
\multicolumn{5}{l}{\scriptsize $^a$ Redshift reference : Struble \& Rood (1999)}\\
\multicolumn{5}{l}{\scriptsize $^b$ X-ray Luminosities reference : Ebeling et al. (1996)}\\
\multicolumn{5}{l}{\scriptsize $^c$ Temperatures reference : Wu et al. (1999)}\\
\multicolumn{5}{l}{\scriptsize $^d$ SZ effect references : Herbig et al. (1995), Coma; Andreani et al. (1996), A2744;}\\
\multicolumn{5}{l}{\scriptsize ~~White \& Silk (1980), A2319.}\\
\label{tablog}
\end{tabular}
\end{flushleft}
\end{table*}

We performed a comparison of the radio and X-ray surface
brightness both using `point' values and annuli-averaged values.
From the first we derive the $F_{Radio}$-$F_X$ correlation. 
The image processing was done using the Synage++ program
(Murgia 2000, in preparation and PhD thesis), which allows to
create grids and concentric rings with the possibility of
excluding areas containing contaminating emission (i.e. from radio
galaxies). For Coma and A2255 we used radio images where discrete
sources were subtracted using the AIPS package, while in the other two
clusters discrete radio sources were excluded from the
analysis with the help of the Synage++ program.

For each cluster, we constructed a square grid covering the region
containing the radio halo (see Fig. \ref{synage} upper panel).  We
derived the mean and the root-mean-square (RMS) of the radio and
X-ray surface brightness for every grid cell. The grid cell
sizes were chosen to obtain a good compromise between high signal
to noise ratio and good resolution.  The RMS is assumed to be an
estimate of the error, and it is used as a weighting factor in the
following fitting procedure. The resulting point-to-point radio versus
X-ray brightness comparisons of the four clusters are presented in
Figs. \ref{A2255_grid90}, \ref{Coma_grid90}, \ref{A2744_grid20},
and \ref{A2319_grid20}. There, each point represents the mean
brightness obtained of each cell of the grid while the error-bars
indicate the RMS of the brightness distribution. Furthermore, the
horizontal dashed lines indicate the 3 sigma noise
level of the radio maps. For graphical reasons the lower error-bars
of radio fluxes below the 3 sigma radio noise levels are not
displayed. We fit the data for a power-law relation
\begin{equation}
\label{eq:powerlaw}
\frac{F_{\rm Radio}}{\rm mJy\, arcsec^{-2}} = a\,\left( \frac{F_{\rm
X}}{\rm mCounts\,s^{-1}\,arcsec^{-2}} \right)^b,
\end{equation}
using logarithmic units and using error-bars for both $F_{\rm
Radio}$ and $F_{\rm X}$. The best fit parameters for each cluster
were estimated by a least square method and the corresponding fitting
line is indicated by a solid line in each figure.  A larger grid
cell does not change the fitting results.
However, since this approach of counts in cells involves 
spatial averaging, this kind of analysis is insensitive to 
scale features smaller than the cell size.

Due to some degree of morphological difference between the
 radio and X-ray images deviations from a tight correlation
exist, which appear as scatter in the $F_{Radio}$-$F_X$ plots.
However, most of the outlayers have radio fluxes below the 3
sigma noise level.

We also compared the shape of the normalized radio and X-ray
brightness profiles of the clusters in Fig. \ref{A2255_prof},
\ref{Coma_prof}, \ref{A2744_prof}, and \ref{A2319_prof}.  To obtain
these profiles, the mean and RMS brightnesses were estimated within
concentric rings centered on the X-ray brightness peaks (see
Fig. \ref{synage} lower panel).  The RMS within the rings was used
as an estimate of the errors.

As can be seen from the large error-bars in the profile, but also
from the brightness variation
within the concentric rings (see lower panel of Fig. \ref{synage}),
morphological distortions of the clusters do affect the profiles.
Moreover, since the X-ray and the radio peaks are not always perfectly
coincident, also the choice
of the center of the concentric rings could introduce some arbitrariness
into the analysis.
For these reasons, the local point-to-point comparison
is more reliable in the analysis of radio halos.

\section{Results}

\subsection{A2255}

\begin{figure}
\resizebox{9 cm}{!}{\includegraphics {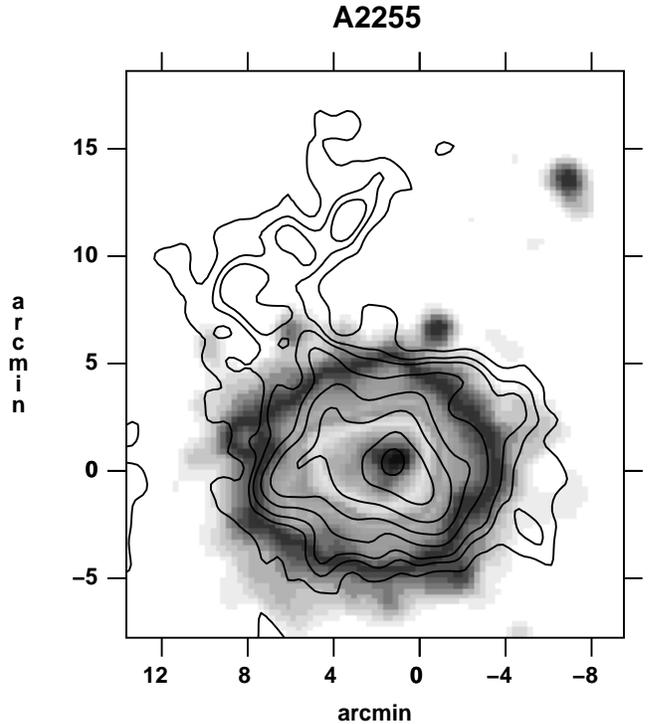}}
\hfill
\caption[]{Contours of the halo image at 90 cm, after subtraction
of discrete sources, superimposed on the grey-scale X-ray PSPC image
of the cluster A2255.  Contour levels are: 3, 6, 8, 12, 17, 24, 34, 40
mJy/beam.  A similar image was previously published in Feretti et
al. (1997a).}
\label{syn_A2255}
\end{figure}

An X-Ray study of the cluster A2255, located at redshift 0.0806
(Struble \& Rood 1999), has been presented by Burns et al. (1995) and
Feretti et al. (1997a).  The cluster shows an asymmetric X-ray
structure suggesting an ongoing cluster merger event.

In Fig. \ref{syn_A2255} we show the overlay between the radio
(contours) and the X-ray surface brightness image (grey scale).  
The X-ray image results
from 14500 seconds exposure time of the Rosat PSPC detector. It was
obtained by binning the photon events in a two dimensional grid and
then smoothing with a Gaussian of $\sigma$=30$''$.

A high sensitivity radio map was presented by Feretti et
al. (1997a). The image was obtained with the Westerbork Synthesis
Radio Telescope (WSRT) at 90 cm and with a resolution of
88$''$ $\times$ 82$''$. The cluster radio emission is characterized by a
diffuse halo source located at the cluster center, by a peripheral
relic in the North-East, and by the presence of several tailed
radio galaxies.  The radio galaxies have been subtracted from the
image used here (Feretti et al. 1997a, Fig. \ref{synage} upper panel
and Fig. \ref{syn_A2255}). The angular size of the halo is about
10$'$.

\begin{figure}
\resizebox{9 cm}{!}{\includegraphics {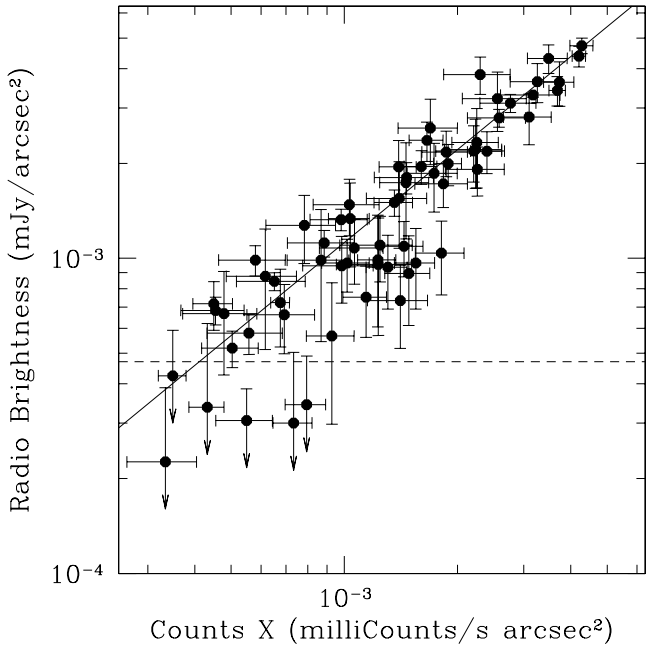}}
\hfill
\caption[]{$F_{\rm Radio}-F_{\rm X}$ relation of the A2255 cluster.
Here and in the following figures the points show the mean of the
brightness obtained within each cell of the grid while the error-bars
indicate the RMS of the brightness distribution. The horizontal dashed
line indicates the 3 sigma noise level of the radio map. For graphical
reasons we indicate with an upper limit the data points below the 3
sigma radio noise level.  The best fit is indicated by a solid line.
}
\label{A2255_grid90}
\end{figure}

\begin{figure}
\resizebox{9 cm}{!}{\includegraphics {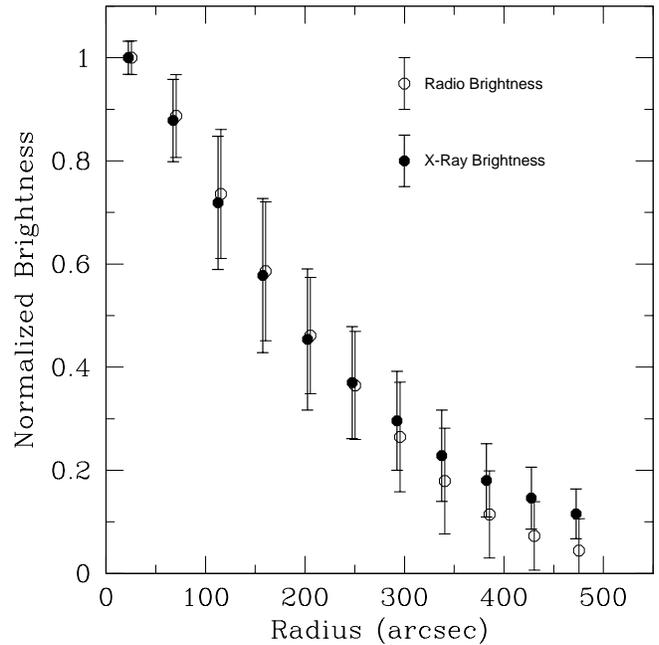}}
\hfill
\caption[]{Comparison between the normalized radio and the X-Ray
brightness profiles in A2255. The radio data points are offset by
3$''$ for clarity.}
\label{A2255_prof}
\end{figure}

In Fig. \ref{synage} we show the grid and rings for the
analysis of this cluster: the grid cells have a size of 90$''$ (184
kpc), the width of the concentric rings is 45$''$.
Fig. \ref{A2255_grid90} and Fig. \ref{A2255_prof} show the two
comparisons of the radio and X-ray emissions.

Fig. \ref{A2255_grid90} indicates that the radio brightness correlates
well with the X-ray brightness.  The normalization and slope of the
fitted power-law (Eq. \ref{eq:powerlaw}) are $a =0.97 \pm 0.25$,
and $b=0.98 \pm 0.04$. The formal errors result from the RMS of
the grid data-points. The derived correlation is consistent with a
linear relation between the radio and X-ray emissivity.

Fig. \ref{A2255_prof} shows that the two brightness profiles are
very similar.

\subsection{Coma (Abell 1656)}

The radio halo in Coma (Coma C) is the prototype
and the best studied example of cluster radio halos.
It is located at the cluster center, shows a rather regular shape and a
large size.

\begin{figure}
\resizebox{9 cm}{!}{\includegraphics {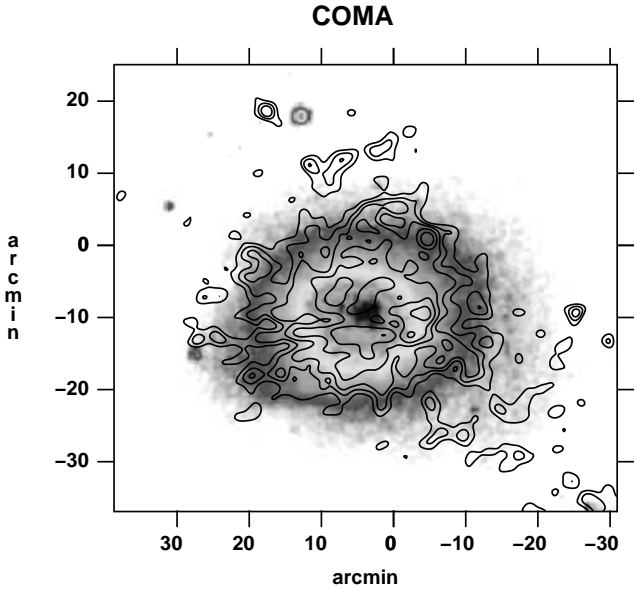}}
\hfill
\caption[]{Contours of the halo image at 90 cm, after subtraction
of discrete sources, superimposed on the grey-scale X-ray PSPC image
of the cluster Coma. 
Contour levels are: 5, 8, 11, 16, 23, 32 mJy/beam.
The radio image is published in Feretti (2000),
while the X-ray image is from White et al. (1993).}
\label{syn_Coma}
\end{figure}

Deiss et al. (1997) obtained a map at 1.4 GHz of the halo, after
subtraction of all discrete sources. They pointed out the similar
extension of the X-ray and the radio emission in E-W direction towards
the NGC 4839 group.  According to Burns et al. (1994) the X-ray
morphology indicates that the Coma cluster has undergone a collision
with the NGC 4839 group about 2 Gyr ago.  Moreover, recent works
provide evidence for ongoing merging of groups in the center of
Coma (Colless \& Dunn 1996, Vikhlinin et al. 1997, Donnelly et
al. 1999, Bravo-Alfaro et al. 2000).

We use here the radio map of Coma obtained with the Westerbork
Synthesis Radio Telescope at 90 cm at an angular resolution of
125$''$$\times$55$''$ (Feretti \& Giovannini 1998) after subtraction
of the discrete sources. In this radio image the projected size
of the halo is about 30$'$. 
The X-ray image was obtained from Rosat PSPC observations
(White et al. 1993) and was kindly supplied by Dr. Briel.

In Fig. \ref{syn_Coma} we show the overlay between the radio
(contours) and the X-ray surface brightness (grey scale) images.
For a better display (but not for the data analysis) the radio map was
smoothed, whith a Gaussian of 125$''$$\times$125$''$ (FWHM).

\begin{figure}
\resizebox{9 cm}{!}{\includegraphics {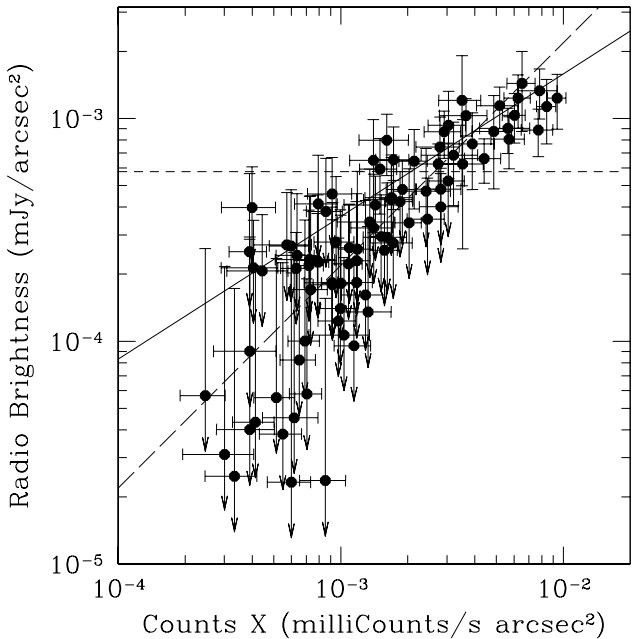}}
\hfill
\caption[]{$F_{\rm Radio}$ vs. $F_X$ for Coma, as in
Fig. \ref{A2255_grid90}. The best fit is indicated by a solid line,
while we show with the long dashed line the linear relation
(line with $b=1$) between the radio and the X-Ray Brightness.}
\label{Coma_grid90}
\end{figure}

\begin{figure}
\resizebox{9 cm}{!}{\includegraphics {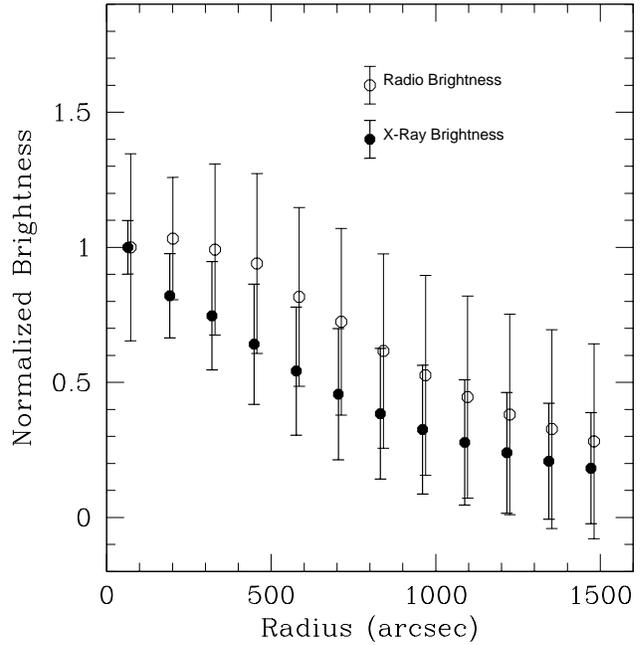}}
\hfill
\caption[]{Comparison between the normalized
radio and the X-Ray brightness profiles in Coma.
The radio data points are offset by
9$''$ for clarity.
}
\label{Coma_prof}
\end{figure}

The grid cells have a size of 256$''$ (166 kpc), while the concentric
rings have a thickness of 128$''$.  The shape of the two radial profiles
are slightly different and the point-to-point comparison of radio and
X-ray brightness shows a large scatter, which can also be seen in the
radial profiles. It might be worth to emphasize that the radio
brightness emission in Coma is lower than in the case of the other
analyzed clusters: only a minority of the data points are
above the 3 sigma noise level (Fig. \ref{Coma_grid90}). This possibly
allows background noise to introduce additional
scatter. However, the error-weighting of the data should reduce the
noise influence, although it could introduce a bias towards flatter $F_{\rm
Radio}-F_{\rm X}$ relations (i.e. smaller values of $b$). The fitted
power-law relation (Eq. \ref{eq:powerlaw}) has the normalization $a
=0.03 \pm 0.01$ and slope $b = 0.64 \pm 0.07$.  We emphasize that
this sub-linear power law relation should be verified through a
radio map of higher sensitivity.

Fig. \ref{Coma_grid90} shows that a linear relation ($b=1$)
between grid values of the radio and X-Ray brightness may
also be consistent with the data.

\subsection{A2744}

The radio halo in A2744 was found by Giovannini et al. (1999) 
during a search for new radio halos and cluster
relics among the candidates extracted from the NRAO VLA Sky
Survey (NVSS, Condon et al. 1998).  The cluster A2744, located at a
redshift of 0.308 (Struble \& Rood 1999), was re-observed with the
VLA at 1.4 GHz in C and D configuration (Govoni et al. 2000, in
preparation).  We analyzed the resulting map, which has an angular
resolution of 50$''$ $\times$ 50$''$.

\begin{figure}
\resizebox{9 cm}{!}{\includegraphics {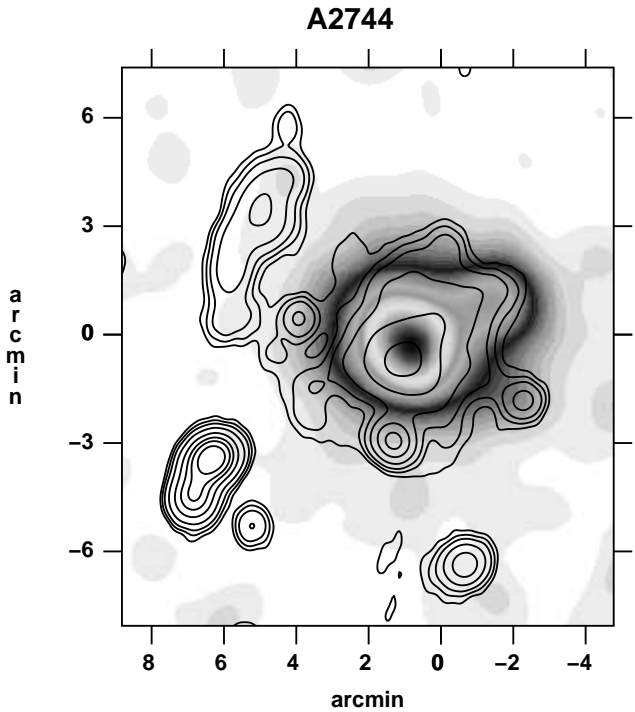}}
\hfill
\caption[]{Contours of the halo image at 20 cm, superimposed on the
grey-scale X-ray PSPC image
of the cluster A2744.
Contour levels are: 0.3, 0.5, 0.8, 1.5, 3, 5, 7, 10, 15,
20, 25, 30, 50 mJy/beam.}
\label{syn_A2744}
\end{figure}

The X-ray image results from 13600 seconds exposure time and was
obtained from Rosat PSPC archive data, by binning the photon events in
a two dimensional grid and then smoothing with a Gaussian function
with $\sigma$=30$''$.  The X-ray background, of about $1.0\times
10^{-7}$ cts sec$^{-1}$ arcsec$^{-2}$, was determined by fitting 
the X-ray surface brightness profile with a hydrostatic
isothermal model (Cavaliere \& Fusco-Femiano, 1981).

The X-ray image shows morphological substructures, which are
possible indication of a recent merger.

In Fig. \ref{syn_A2744} we present the overlay between the radio
(contours) and the X-ray image (grey-scale) of A2744. The
elongated diffuse radio emission in the peripheral North-East
region of the cluster is classified as a cluster relic, and,
therefore, excluded from the analysis.

We performed the analysis of the radio and the X-ray morphology using
a grid with a cell size of 51$''$ (285 kpc), and concentric rings with a
thickness of 30$''$. Since this cluster is at a large distance, the
gridding corresponds to cells of large linear size.  
However, both the X-ray emission and the radio halo are well 
resolved (FWHM of X-ray and radio profiles $\simeq$ 170$''$
from Fig. 10). Therefore, although the number of statistically 
independent cells and rings in the cluster is lower than in other
clusters, it allows us to perform a quantitative analysis.

The absence of X-ray point sources with an optical or 
radio counterpart doesn't allow us to check if 
the offset between the X-ray and the radio peaks is real or not. 
We note that the offset ($\simeq$ 30$''$) is within the
cell size (51$''$).

\begin{figure}
\resizebox{9 cm}{!}{\includegraphics {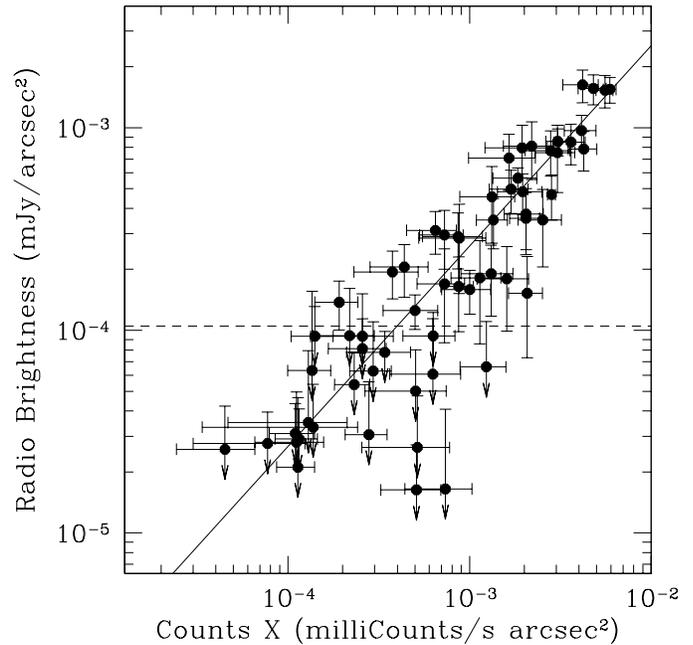}}
\hfill
\caption[]{$F_{\rm Radio}$ versus $F_X$  for A2744, as in Fig. \ref{A2255_grid90}.}
\label{A2744_grid20}
\end{figure}

\begin{figure}
\resizebox{9 cm}{!}{\includegraphics {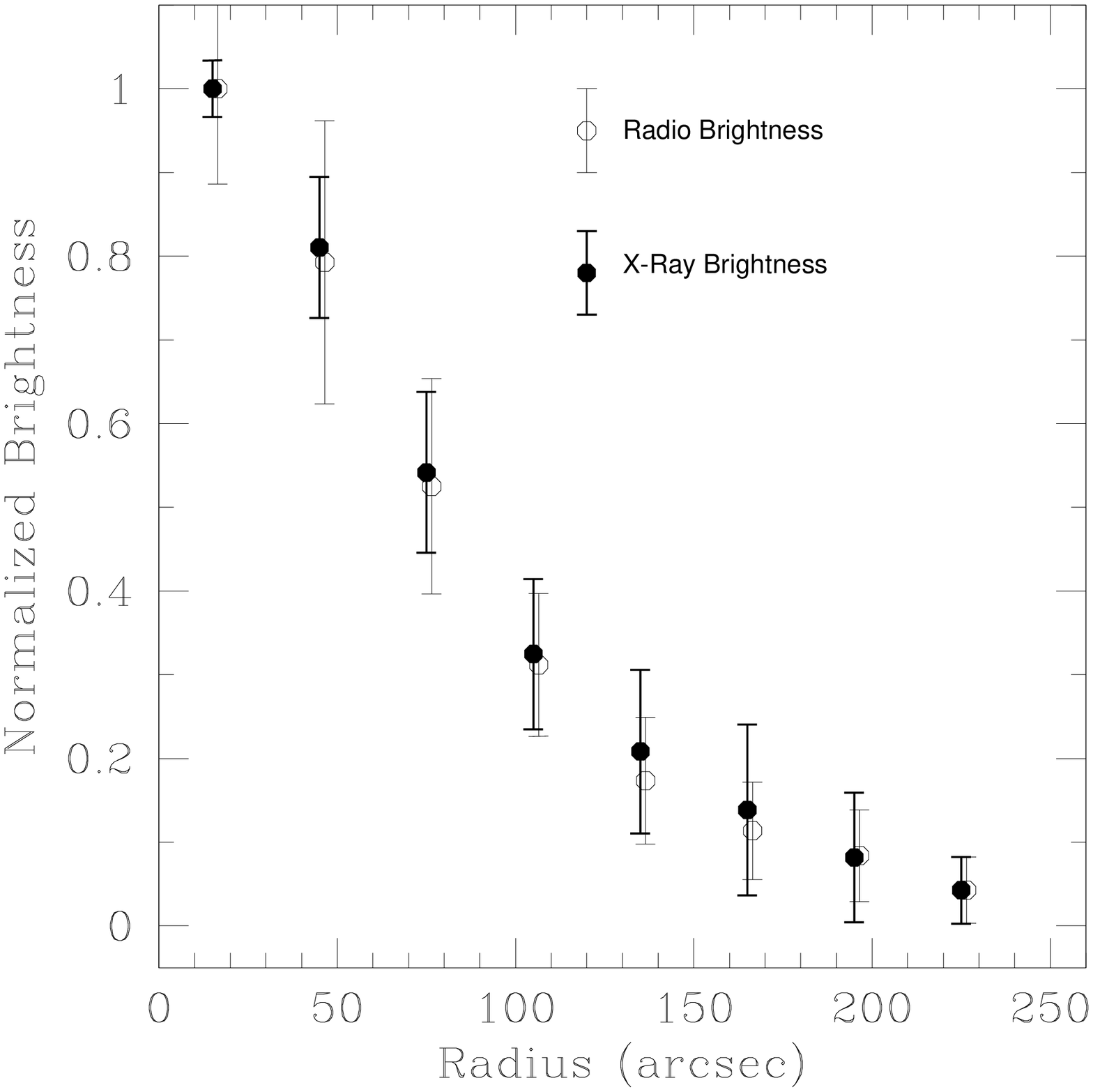}}
\hfill
\caption[]{Comparison between the normalized
radio and the X-Ray brightness profiles in A2744.
The radio data points are offset by
1.5$''$ for clarity.}
\label{A2744_prof}
\end{figure}

Results are presented
in Fig. \ref{A2744_grid20} and \ref{A2744_prof}.
The shape of the two brightness profiles is perfectly coincident
and in this case we get a linear $F_{\rm Radio}-F_X$
relation ($a= 0.24 \pm 0.07$, $b = 0.99 \pm 0.05$).

\subsection{A2319}

\begin{figure}
\resizebox{9 cm}{!}{\includegraphics {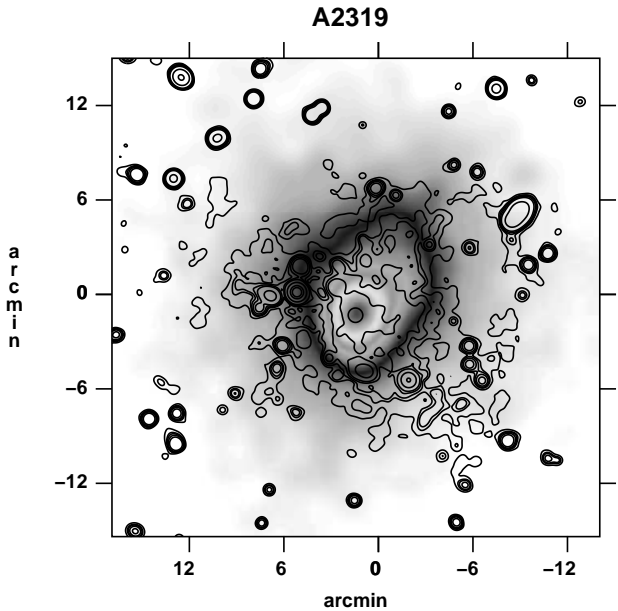}}
\hfill
\caption[]{Contours of the halo image at 20 cm, superimposed on
the grey-scale X-ray PSPC image of the cluster A2319.  Contour levels
are: 0.15, 0.3, 0.4, 0.6, 1, 5, 25, 50 mJy/beam.  A Radio-X ray overlay
of the cluster A2319 was previously published in Feretti et
al. (1997b).}
\label{syn_A2319}
\end{figure}

A radio and X-ray study of the cluster A2319 has been presented by
Feretti et al. (1997b). We refer to this paper for the details of
the images. This cluster is located at a redshift of 0.0557
(Struble \& Rood 1999). It has an irregular radio halo,
with a size of about 16$'$, at its center.

The radio image was
obtained with the Westerbork Synthesis Radio Telescope (WSRT) at 20 cm
with a resolution of 29.0$''$$\times$ 20.4$''$. The X-ray image results
from the Rosat PSPC archive data and was obtained by binning the
photon events in a two dimensional grid and then smoothing with a
Gaussian function with $\sigma$=50$''$.

In Fig. \ref{syn_A2319} we show the overlay between the radio
(contours) and the X-ray surface brightness (grey scale) image. 
For a better display (but not for the data analysis) the radio map was
smoothed, with a Gaussian of 30$''$$\times$30$''$ (FWHM).

\begin{figure}
\resizebox{9 cm}{!}{\includegraphics {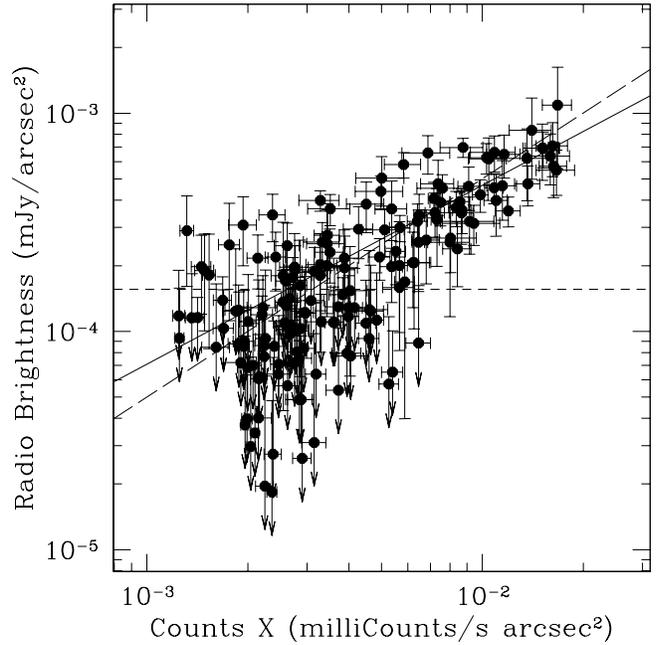}}
\hfill
\caption[]{$F_{Radio}~versus~F_X$ for the A2319 cluster. See
Fig. \ref{A2255_grid90} for details. The best power-law fit
of the radio and the X-ray brightness relation is indicated by a
solid line, and the linear relation ($b=1$) is marked
with a long dashed line.}
\label{A2319_grid20}
\end{figure}

\begin{figure}
\resizebox{9 cm}{!}{\includegraphics {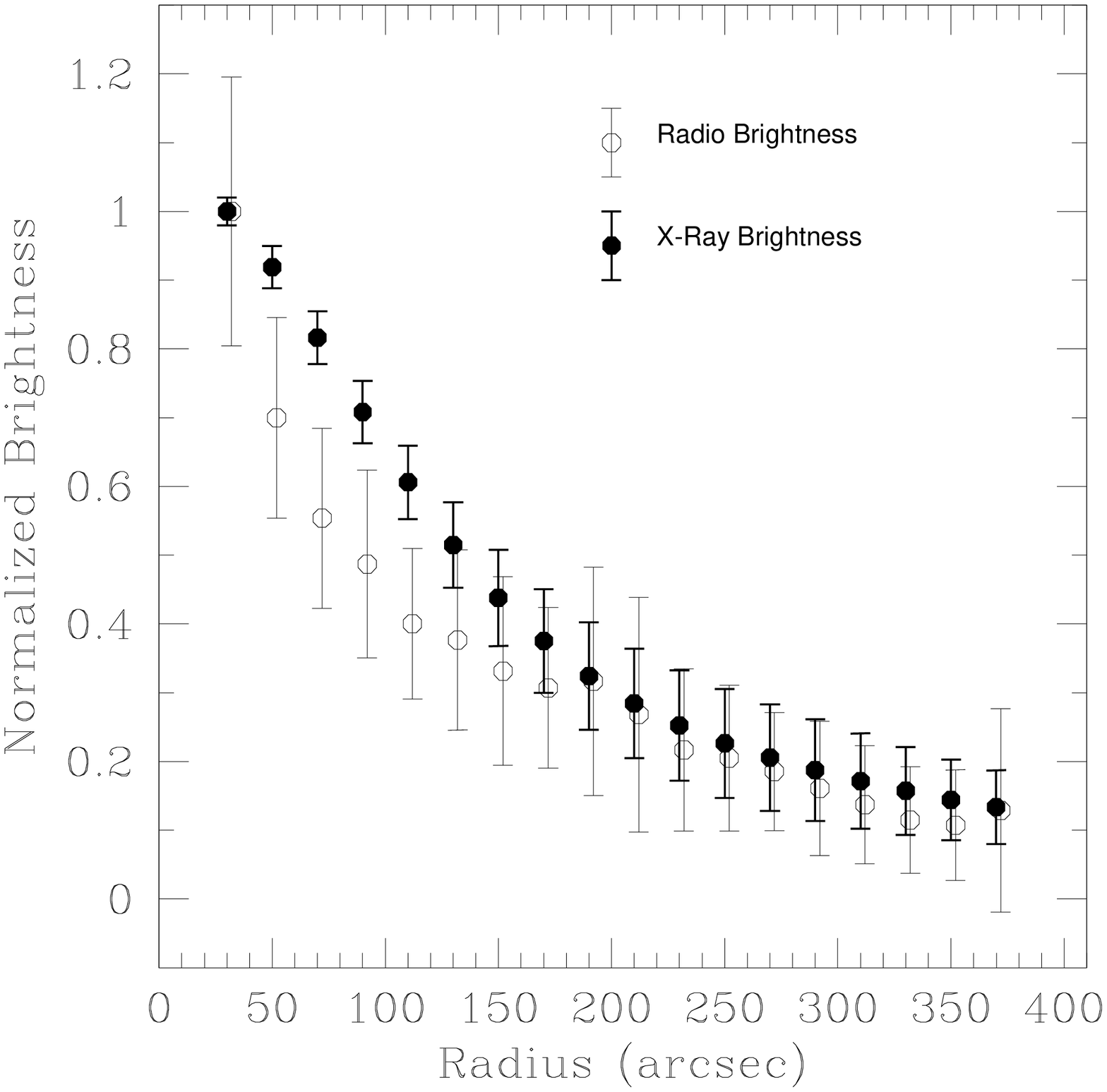}}
\hfill
\caption[]{Comparison between the normalized radio and the X-Ray
brightness profiles in A2319. The radio data points are offset
by 2$''$} for clarity.
\label{A2319_prof}
\end{figure}

The grid has a cell size of 56$''$ (82 kpc), and the concentric rings
have a thickness of 20$''$. Results are given in Fig. \ref{A2319_grid20}
and \ref{A2319_prof}. We get a sub-linear power law $F_{\rm
Radio}$-$F_{\rm X}$ relation with $a= 0.020 \pm 0.004$ and $b= 0.82
\pm 0.04$ (see Fig. \ref{A2319_grid20}). As in the case of
the Coma cluster, the low radio signal to noise ratio might
affect our estimate. The radial brightness profiles have
slightly different shapes.  As for Coma, this could result from
a non-linear $F_{Radio}-F_X$ relation.

\section{Discussion}

The $F_{Radio}-F_X$ point to point comparison studied in this work
gives important insight on the spatial distribution of
magnetic fields and relativistic particles with respect to the
X-ray emitting gas in clusters of galaxies.

We have found a strong positive correlation between the radio
halo and the X-ray brightness in all the analyzed clusters of
galaxies (see Fig. \ref{A2255_grid90}, \ref{Coma_grid90},
\ref{A2744_grid20} and \ref{A2319_grid20}). Moreover, we found a
linear relation between the radio and the X-ray brightness in two
clusters of galaxies (A2255, A2744).

The appearance of radio halos seems to be correlated to the occurrence
of cluster mergers (Feretti et al. 1997b), which suggests
that cluster mergers could be the key of the connection between the
thermal and the relativistic components in the clusters.

In the following we demonstrate how some simple assumptions can
lead to a linear relation between the radio and the X-ray brightness.
Most of the thermal energy content of a galaxy cluster results from
the dissipation of the kinetic energy of gas falling into the cluster
potential. The dissipation occurs in shock waves and turbulent
cascades, which are strongest in the case of major cluster merger
events. Some small fraction of the dissipated energy might go into the
amplification of magnetic fields due to field line stretching in shear
flows (Roettiger et al. 1999, Dolag et at. 1999).  Another fraction
might go into relativistic particle populations via Fermi I or II
acceleration. The rest of the dissipated energy is thermalized in the
hot ICM.  Let us assume for simplicity, that the energy fraction which
goes into the non-thermal components (fields and relativistic
particles) compared to the thermal component is independent of the
position in a cluster.  This assumption needs not to hold in reality,
but it allows us to construct simple test models, for comparison with
observations.

The X-ray emissivity in a cluster is given by:
\begin{equation}
j_X\propto n_\e^2\, (kT_\e)^{1/2},
\label{xemission}
\end{equation}
where $n_\e$ and $kT_\e$ are the electron density and temperature.
For an isothermal cluster the X-ray emissivity at a given position
is proportional to the thermal energy density
($\eps_\th=3\,n_e\,kT_e$) squared:
\begin{equation}
j_X\propto\eps_\th^2\, (kT_\e)^{-3/2}.
\end{equation}

If the number density of relativistic electrons with energy
between $\eps$ and $\eps+d\eps$ can be assumed to be
$N(\eps)d\eps=N_0\eps^{-\delta}d\eps$ the radio emissivity at a given
position in a cluster is given by:
\begin{equation}
 j_{Radio}\propto N_0\, B^{(\delta +1)/2}\, \nu ^{-(\delta-1)/2},
\end{equation}
where $\nu$ is the frequency and $B$ is the magnetic field strength.

In general, for each value of $\delta$, the observationally
obtained linear relation $F_{Radio}\propto F_X$ implies that the
relativistic particles and the magnetic field are connected to the
thermal plasma through the relation:
\begin{equation}
 N_0\, B^{(\delta +1)/2}\,\propto n_\e^2\, (kT_\e)^{1/2}.
\label{fine}
\end{equation}

Such a relation must be approximately fulfilled in A2255 and
A2744. In the other two clusters, the relations obtained between
$F_{Radio}$ and $F_X$ seem to imply that the radial
decline of the magnetic field strength and relativistic
particle density is slower than the decline of thermal gas
density.

In the following we assume for simplicity a single radio
spectral index ($\alpha=\frac{\delta-1}{2}=+1$)
which is typical for cluster radio halos.
Therefore the radio emissivity at frequency $\nu$ is
\begin{equation}
 j_{Radio}\propto \eps_\cre \,\eps_B \,\nu^{-1},
\end{equation}
where $\eps_\cre=\int_{}^{}{\eps N(\eps)d\eps}$
is the relativistic
electron energy density, and $\eps_B = B^2/(8\pi)$ is the
local magnetic field ($B$) energy density.

If we assume -- as suggested above -- that magnetic fields and radio
emitting electrons have energy densities proportional to the thermal
energy content ($\eps_B \propto \eps_\th$ and $\eps_\cre \propto
\eps_\th$) we find that the ratio $j_{Radio}/j_X \propto kT_\e^{3/2}$
is independent of the position within an isothermal cluster. The same
is true for the ratio of the surface brightnesses $F_{Radio}$ and
$F_X$, which are the line-of-sight projected emissivities. Thus
$F_{Radio} \propto F_X$ as observed at least in two of our
cases. Such a linear relation might be obtained by some of the
primary electron models where the energy density in relativistic
electrons is proportional to the thermal energy density.

In hadronic secondary electron models the radio emitting
electrons are secondary particles produced in hadronic cosmic ray
proton-gas interactions (see review En{\ss}lin 2000). In these
models the radio emissivity is given by (Dolag \& En{\ss}lin
2000):
\begin{equation}
 j_{Radio}  \propto \eps_\crp\,n_\e \, \eps_B/(\eps_B + \eps_{\rm
cmb})\, \nu^{-1},
\end{equation}
assuming $\alpha =1$. This more complicated dependence results from
the fact that the spectrum of the radio emitting electrons is
shaped by the parent cosmic ray proton spectrum, here assumed
to be a power-law, and by the electron cooling
processes. These are mainly due to synchrotron losses
($\propto \eps_B$) and to inverse Compton scattering with the
CMB photon field ($\propto \eps_{\rm cmb}$).  We assume further that
the cosmic ray proton population has an energy density
proportional to that of the thermal component ($\eps_\crp
\propto \eps_\th$). The resulting ratio of the radio and X-ray
emissivities
\begin{equation}
 j_{\rm Radio} /j_X\propto kT^{1/2} \, \eps_B/(\eps_B + \eps_{\rm cmb})
\end{equation}
is independent of the magnetic energy density for strong fields
($\eps_B \gg \eps_{\rm cmb}$), and proportional to it for weak fields
($\eps_B \ll\eps_{\rm cmb}$). Thus, for a radial decreasing
magnetic field strength we expect a linear relation between the radio
and the X-ray brightness in the case of strong magnetic fields, while
we expect a super-linear power law in the case of weak magnetic
fields. For instance, for a magneto-hydrodynamically simulated
Coma-like cluster of galaxies with intermediate field strength
($\eps_B \ls\eps_{\rm cmb}$), which is consistent with published
Faraday rotation measurements, Dolag \& En{\ss}lin (2000) get a radio
to X-ray surface brightness relation of $F_{\rm Radio}\sim
F_X^{1.26}$. The prediction of this specific version of hadronic
secondary electron models is in contradiction with the data
analysis presented here.

\section{Conclusions\label{sec:discussion}}

We analyzed radio and X-ray images of four clusters of galaxies on a
point-to-point and also on a circular averaged (radial profile)
basis. We argue that the point-to-point comparison is less affected by
morphological distortions, and that it allows to introduce a radio vs
X-ray surface brightness relation ($F_{\rm Radio}$-$F_{\rm X}$
relation). This was found to be consistent with a linear relation in
two clusters (Abell 2555: $b = 0.98 \pm 0.04$; Abell 2744: $b=
0.99\pm0.05$) and to be a sub-linear power law in the other two
analyzed clusters (Coma: $b= 0.64\pm0.07$; Abell 2319: $b= 0.82 \pm
0.04$). We note that in the case of Coma the relation could be
affected by the low signal to noise ratio of the used radio map.

We presented two simplified cluster radio halo models: a primary
electron and a hadronic secondary electron model. In both models we
assumed that the magnetic fields and the primary particle population
(electrons in the primary model, the parent protons in the hadronic
secondary model) have energy densities proportional to the thermal
energy density. The predicted $F_{\rm Radio}$-$F_{\rm X}$ relations of
these models are linear for the primary model and for the hadronic
secondary model in the regime of strong magnetic fields ($\eps_B \gg
\eps_{\rm cmb}$). The hadronic secondary model with weak or moderate
field strength ($\eps_B \ls\eps_{\rm cmb}$) and with an assumed linear
scalings of the thermal and non-thermal (relativistic electrons and
magnetic fields) energy densities produces a super-linear power law
$F_{\rm Radio}$-$F_{\rm X}$ relation, and it is therefore disfavored
by the data analysis presented here.

We hope that the point-to-point radio vs X-ray surface brightness
relation will prove a useful observational and theoretical tool for
future studies of the mysterious nature of cluster radio halos and of
the non-thermal energy content of clusters of galaxies.\\

\section*{Acknowledgments}
We are indebted to Matteo Murgia for the use of the Synage++ program
and for helpful comments on the paper.  We thank Ulrich G. Briel who
kindly supplied the X-Ray map of Coma.  We further thank Gianfranco
Brunetti for helpful discussions and Francesco Miniati for comments
on the manuscript.

\end{document}